\documentclass[12pt]{article}

\usepackage{latexsym}
\usepackage{amssymb}
\usepackage{array}

\def\0{\over } \def\1{\vec } \def\2{{1\over2}} \def\4{{1\over4}}
\def\5{\bar } 
\def\6{\partial }
\def\7#1{{#1}\llap{/}}
\def\8#1{{\textstyle{#1}}} \def\9#1{{\bf{#1}}}
\def\.{\cdot }
\def\^#1{\widehat{#1}}
 
  \let\g=\gamma 
   
  \let\l=\lambda \let\m=\mu
    
\let\t=\tau  
\let\ph=\varphi

\def\CL{{\cal L}}

\def\({\left(} \def\){\right)} \def\<{\langle } \def\>{\rangle }
\def\[{\left[} \def\]{\right]}  
 
\def\pmbf#1{\setbox0=\hbox{${#1}$}
        \kern-.025em\copy0\kern-\wd0
        \kern.05em\copy0\kern-\wd0
        \kern-.025em\raise.0433em\box0 }

\def\be{\begin{equation}}
\def\ee{\end{equation}}

\newcommand{\bel}[1]{\begin{equation}\label{#1}}
\def\bea{\begin{eqnarray}}
\newcommand{\beal}[1]{\begin{eqnarray}\label{#1}}
\def\eea{\end{eqnarray}}

\newcommand{\ds}{\!\not\!\partial}

\newcommand{\intsum}{\sum\!\!\!\!\!\!\!\int}  

\begin{document}

\begin{titlepage}
\renewcommand{\thefootnote}{\alph{footnote}}
~\vspace{-2cm}
\begin{flushright} 
TUW-02-16\\
YITP-02-35
\end{flushright}  
\vfil
\centerline{\large\bf The anomaly in the central charge of the supersymmetric kink}
\medskip
\centerline{\large\bf from dimensional regularization and reduction}
\begin{center}
\vfil 
{\large  
A. Rebhan$^1$\footnote{\footnotesize\tt rebhana@hep.itp.tuwien.ac.at}, 
P. van Nieuwenhuizen$^2$\footnote{\footnotesize\tt vannieu@insti.physics.sunysb.edu} 
and R. Wimmer$^1$\footnote{\footnotesize\tt rwimmer@hep.itp.tuwien.ac.at}
}\\
\end{center}  \medskip \smallskip \qquad \qquad 
{\sl $^1$} \parbox[t]{12cm}{ \sl 
  Institut f\"ur Theoretische Physik, Technische Universit\"at Wien, \\
  Wiedner Hauptstr. 8--10, A-1040 Vienna, Austria\\ } \\
\bigskip \qquad \qquad 
{\sl $^2$} \parbox[t]{12cm}{ \sl 
  C.N.Yang Institute for Theoretical Physics, \\
  SUNY at Stony Brook, Stony Brook, NY 11794-3840, USA\\ } \\
\vfil
\centerline{ABSTRACT}\vspace{.5cm}
We show that the anomalous contribution to the central charge
of the 1+1-dimen\-sion\-al $N$=1 supersymmetric kink that is required for
BPS saturation at the quantum level can be linked to an
analogous term in the extra momentum operator of a
2+1-dimensional kink domain wall with spontaneous parity violation
and chiral domain wall fermions. In the quantization of the
domain wall, BPS saturation is preserved by nonvanishing quantum
corrections to the momentum
density in the extra space dimension. Dimensional reduction
from 2+1 to 1+1 dimensions preserves the unbroken $N$=$1\over2$ supersymmetry
and turns these parity-violating contributions into
the anomaly of the central charge of the
supersymmetric kink.
On the other hand, standard dimensional regularization by dimensional
reduction from 1 to $(1-\epsilon)$ spatial dimensions, which also
preserves supersymmetry, obtains the anomaly from an evanescent counterterm.

\end{titlepage}

\setcounter{footnote}{0}

\section{Introduction}

The calculation of quantum corrections to the mass of
a supersymmetric (susy) kink 
and to its central charge 
has proved to be a highly nontrivial
task fraught with subtleties and pitfalls.

Initially it was thought that supersymmetry would
lead to a complete cancellation of quantum corrections \cite{D'Adda}
and thereby guarantee Bogomolnyi-Prasad-Sommerfield
(BPS) saturation at the quantum level.
Then, by considering a kink-antikink system in a finite
box and regularizing the ultraviolet divergences by
a cutoff in the number of the discretized modes, 
Schonfeld \cite{Schonfeld:1979hg}
found that there is a nonzero, negative
quantum correction at one-loop level, $\Delta M^{(1)}=-m/(2\pi)$,
but remarked that 
``the familiar sum of frequencies \ldots is unacceptably sensitive
to the definition of the infinite volume limit''.
Most of the subsequent literature
\cite{Kaul:1983yt,Chatterjee,Yamagishi:1984zv} 
considered instead a single kink directly, using 
an energy-momentum cutoff which gave
again a null result. A direct calculation of the
central charge \cite{Imbimbo:1984nq} also gave a null result, apparently
confirming a conjecture\footnote{``While 
we suspect that this is true we have no
proof.''  \cite{Witten:1978mh}}
of Witten and Olive \cite{Witten:1978mh}
that BPS saturation in the minimally susy 1+1 dimensional case
would hold although arguments on multiplet shortening do not
seem to apply.

In Ref.~\cite{Rebhan:1997iv} two of the present authors
noticed a surprising dependence on the regularization method,
even after the renormalization scheme has been fully fixed.
In particular it was found that the
naive energy-momentum cutoff
as used in the susy case spoils the integrability of
the bosonic sine-Gordon model \cite{Dashen:1975hd}. Using a mode regularization
scheme and periodic boundary conditions in a finite box
instead led to a susy kink mass correction
$\Delta M^{(1)}=+m(1/4-1/2\pi)>0$ (obtained previously also in
Ref.~\cite{Uchiyama:1986gf})
which together with
the null result for the central charge appeared to
be consistent with the BPS bound, but implying nonsaturation.
Subsequently it was found by two of us together with
Nastase and Stephanov \cite{Nastase:1998sy} that the
traditionally used periodic boundary conditions are questionable. Using
instead topological boundary conditions which are invisible
in the topological and in the trivial sector together with 
a ``derivative regularization''\footnote{In 
mode regularization it turns out that one has to average over
sets of boundary conditions to cancel both localized
boundary energy and delocalized momentum 
\cite{Goldhaber:2000ab,Goldhaber:2002mx}.}
indeed led to a different result,
namely that originally obtained by Schonfeld \cite{Schonfeld:1979hg},
which however appeared to be in conflict with the BPS inequality
for a central charge without quantum corrections.

Since this appeared to be a pure one-loop effect,
Ref.~\cite{Nastase:1998sy} proposed ``\ldots the interesting
conjecture that it may be formulated in terms of a
topological quantum anomaly.''
It was
then shown by Shifman et al. \cite{Shifman:1998zy}, using a susy-preserving
higher-derivative regularization method, 
that there is an anomalous contribution to the
central charge balancing the quantum corrections to the
mass so that BPS saturation remains intact. In fact, it was later
understood that multiplet shortening does in fact occur even
in minimally susy 1+1 dimensional theories, giving rise
to single-state supermultiplets \cite{Losev:2001uc,Losev:2000mm}.

Both results, the nonvanishing mass correction and thus the necessity
of a nonvanishing correction to the central charge,
have been confirmed by a number of different methods 
\cite{Graham:1998qq,Litvintsev:2000is,Goldhaber:2000ab,Goldhaber:2001rp,Wimmer:2001yn,Bordag:2002dg,Goldhaber:2002mx}
validating also the finite mass formula in terms of only
the discrete modes derived in 
Refs.~\cite{Casahorran:1989vd,Boya:1990fp} based on the
method of \cite{Cahill:1976im}.
However, some authors claimed a nontrivial quantum correction
to the central charge \cite{Graham:1998qq,Casahorran:1989nr}
apparently without the need of the anomalous term
proposed in Ref.~\cite{Shifman:1998zy}.

In a previous paper \cite{Rebhan:2002uk}, we have shown
that a particularly simple and elegant regularization scheme
that yields the correct quantum mass of the susy kink
is dimensional regularization, if the kink is embedded
in higher dimensions as a domain wall
\cite{Bollini}.
Such a scheme was not considered before for the susy kink
because both susy and the existence of finite-energy
solutions seemed to tie one to one spatial dimension.

In fact, the 1+1 dimensional susy kink can be embedded in
2+1 dimensions with the same field content while keeping
susy invariance. For the corresponding classically
BPS saturated domain wall (a 1+1 dimensional object by itself), 
we have found a
nontrivial negative correction to the surface (i.e. string) tension
\cite{Rebhan:2002uk}.
In order to have BPS saturation at the quantum level, there
has to be a matching correction to the momentum in the extra
dimension which is the analog of the central charge of the 1+1
dimensional case.

In this work we show that in dimensional regularization
by means of dimensional reduction from 2+1 dimensions, which
preserves susy, one finds the required correction
to the extra momentum to have a BPS saturated domain wall
at the quantum level. This nontrivial correction is made
possible by the fact that the 2+1 dimensional theory spontaneously breaks
parity, which also allows the appearance of domain wall fermions
of only one chirality.

By dimensionally reducing to 
1+1 dimensions, the parity-violating contributions
to the extra momentum turn out to provide an
anomalous contribution to the central charge as postulated
in Ref.~\cite{Shifman:1998zy}, thereby giving a novel physical
explanation of the latter. 
This is in line with the well-known fact that central charges of
susy theories can be reinterpreted as momenta in higher dimensions.

Hence, in the case of the susy kink, dimensional regularization is seen
to be compatible with susy invariance only at the expense of a 
spontaneous parity
violation, which in turn allows nonvanishing quantum corrections
to the extra momentum in one higher spatial dimension.
On the other hand, the surface term that usually 
exclusively provides the central charge does not receive quantum 
corrections in dimensional regularization, by the same reason
that led to null results previously in other schemes
\cite{Imbimbo:1984nq,Rebhan:1997iv,Nastase:1998sy}.
The nontrivial anomalous quantum correction to the central charge operator
is thus seen to be entirely the remnant of the spontaneous parity violation
in the higher-dimensional theory in which a susy kink
can be embedded by preserving minimal susy.

We also (in Sect.~\ref{sec33}) pinpoint what we believe to be the
error in Ref.~\cite{Graham:1998qq} who arrived
at the conclusion of BPS saturation apparently without the
need for an anomalous additional term in the central charge operator. 

In the last section, we consider dimensional regularization
by dimensional reduction from 1 to 1-$\epsilon$ spatial
dimensions, which also preserves supersymmetry.
In this case we show that an anomalous contribution to the central
charge arises from the necessity to add an evanescent
counterterm to the susy current. This counterterm preserves susy but
produces an anomaly in the conformal-susy current. 
We also construct the conformal central-charge current whose
divergence is proportional to the
ordinary central-charge current and thus
contains the central-charge anomaly as
superpartner of the conformal-susy anomaly.

\section{Minimally supersymmetric kink and kink domain wall}

\subsection{The model}

The real $\ph^4$ model in 1+1 dimensions with spontaneously
broken $Z_2$ symmetry ($\ph\to-\ph$) has topologically
nontrivial finite-energy solutions called ``kinks'' which
interpolate between the two degenerate vacuum states $\ph=\pm v$.
It has a minimally supersymmetric extension 
\cite{DiVecchia:1977bs}
\bel{Lss}
\CL=-\2\left[ (\6_\m\ph)^2+U(\ph)^2+\5\psi\g^\m\6_\m\psi+
U'(\ph)\5\psi\psi \right]
\ee
where $\psi$ is a Majorana spinor, $\5\psi=\psi^{\mathrm T} C$ with
$C\gamma^\mu=-(\gamma^\mu)^TC$. We shall use
a Majorana representation of the Dirac matrices with $\g^0=-i\t^2$,
$\g^1=\t^3$, and $C=\t^2$ in terms of the
standard Pauli matrices $\tau^k$ so that $\psi={\psi^+\choose\psi^-}$ with
real $\psi^+(x,t)$ and $\psi^-(x,t)$.

The $\ph^4$ model 
is defined as the special case
\be\label{Uphi}
U(\ph)=\sqrt{\l\02}\(\ph^2-v_0^2\),\qquad v_0^2\equiv \mu_0^2/\l
\ee
where the $Z_2$ symmetry of the susy action also involves the fermions
according to $\ph\to-\ph, \psi\to\gamma^5\psi$ with 
$\gamma^5=\gamma^0\gamma^1$.
A classical kink at rest at $x=0$ which interpolates
between the two vacua $\ph=\pm v_0$ is given by \cite{Raj:Sol}
\bel{Ksol}
\ph_{K}=v_0 
\tanh\(\mu_0x/\sqrt2\).
\ee

At the quantum level we have to renormalize, and we shall employ
the simplest possible scheme\footnote{See \cite{Rebhan:2002uk} for
a detailed discussion of more general renormalization schemes
in this context.} which consists of putting all
renormalization constants to unity except for a mass counterterm
chosen such that tadpole diagrams cancel completely in
the trivial vacuum. At the one-loop
level and using dimensional regularization this gives
\be\label{deltamu2}
\delta \mu^2 = \lambda\, \delta v^2 = \lambda
\int{dk_0d^{d}k\0(2\pi)^{d+1}}{-i\0k^2+m^2-i\epsilon}= \lambda
\int{d^{d}k\0(2\pi)^d}{1\02[\vec k^2+m^2]^{1/2}},
\ee
where $m=U'(v)=\sqrt{2}\mu$ is the mass of elementary bosons and
fermions and $k^2=\vec k^2-k_0^2$.

The susy invariance of the model 
under $\delta\varphi=\bar\epsilon\psi$ and $\delta\psi=
(\not\!\partial\ph-U)\epsilon$
(with $\mu_0^2$ replaced by
$\mu^2+\delta\mu^2$) leads to the on-shell conserved Noether
current
\be\label{susyj}
j_\mu=-(\not\!\partial\ph+U(\ph))\gamma_\mu\psi
\ee
and two conserved charges $Q^\pm=\int dx\,j_0^\pm$.

The model (\ref{Lss}) is equally supersymmetric in 2+1 dimensions,
where we use $\gamma^2=\tau^1$. The same renormalization scheme
can be used,
only the renormalization constant (\ref{deltamu2}) has
to be evaluated for $d=2-\epsilon$ in place of $d=1-\epsilon$
spatial dimensions.

While classical kinks in 1+1 dimensions 
have finite energy (rest mass) $M=m^3/\lambda$,
in (noncompact) 2+1 dimensions there exist no longer solitons of
finite-energy. Instead one can have (one-dimensional) domain walls with
a profile given by (\ref{Ksol}) which have finite surface (string) tension
$M/L=m^3/\lambda$. With
a compact extra dimension one can of course use these configurations
to form ``domain strings'' of finite total energy proportional
to the length $L$ of the string when
wrapped around the extra dimension. 

The 2+1 dimensional case is different also with respect to the
discrete symmetries of (\ref{Lss}). In 2+1 dimensions, 
$\gamma^5=\gamma^0\gamma^1\gamma^2 = \pm {\bf 1}$ corresponding
to the two inequivalent choices available for $\gamma^2=\pm\tau^1$
(in odd space-time dimensions the Clifford algebra has
two inequivalent irreducible representations).
Therefore, the sign of the fermion mass (Yukawa) term can no longer
be reversed by $\psi\to\gamma^5\psi$ and there is no longer
the $Z_2$ symmetry  $\ph\to-\ph, \psi\to\gamma^5\psi$.

What the 2+1 dimensional model does break spontaneously is instead
{\em parity}, which corresponds
to changing the sign of one of the spatial coordinates.
The Lagrangian is invariant under $x^m \to -x^m$ for
a given spatial index $m=1,2$ together with $\ph\to-\ph$ (which
thus is a pseudoscalar) and $\psi\to\gamma^m \psi$.
Each of the trivial vacua breaks these invariances spontaneously,
whereas a kink background in the $x^1$-direction with
$\ph_K(-x^1)=-\ph_K(x^1)$ is symmetric with respect to
$x^1$-reflections, but breaks $x^2=y$ reflection invariance.
 
This is to be contrasted with the 1+1 dimensional case, where
parity ($x^1\to-x^1$) can be represented either by $\psi\to\gamma^0\psi$
and a true scalar $\ph\to\ph$ or by $\psi\to\gamma^1\psi$ and
a pseudoscalar $\ph\to-\ph$. The former leaves the trivial vacuum
invariant, and the latter the ground state of the kink sector.

\subsection{Susy algebra}

The susy algebra for the 1+1 and the 2+1 dimensional cases can
both be covered by starting from 2+1 dimensions, the 1+1 dimensional
case following from reduction by one spatial dimension.


In 2+1 dimensions one
obtains classically \cite{Gibbons:1999np}
\begin{eqnarray}
  \label{eq:3dsusy}
 \{Q^{\alpha},\bar{Q}_{\beta}\}&=&2i(\gamma^M)^{\alpha}{}_{\beta}P_M\ ,
          \quad (M=0,1,2)\nonumber\\
    &=&2i(\gamma^0H+\gamma^1(\tilde{P}_x+\tilde{Z_y})
    +\gamma^2(\tilde{P}_y-\tilde{Z}_x))^\alpha{}_\beta,
\end{eqnarray}%
where we separated off two surface terms $\tilde Z_m$ in defining
\begin{eqnarray}
  \label{eq:Ptilde}
\tilde P_m = \int d^dx  \tilde{\mathcal{P}}_m, \quad 
  &&\tilde{\mathcal{P}}_m=\dot\ph\,\partial_m\ph
      -\2(\bar{\psi}\gamma^0\partial_m\psi),\\
  \label{eq:Ztilde}
\tilde Z_m = \int  d^dx  \tilde{\mathcal{Z}}_m, \quad
  &&\tilde{\mathcal{Z}}_m=U(\ph) \partial_m\ph = \partial_m W(\ph)
\end{eqnarray}
with $W(\ph)\equiv\int d\ph\, U(\ph)$.

Having a kink profile in the $x$-direction, which satisfies the
Bogomolnyi equation $\partial_x \ph_K=-U(\ph_K)$, one finds
that with our choice of Dirac matrices
\bea
  \label{eq:qpm}
  &&Q^{\pm}=\int d^2x[(\dot\ph\mp\partial_y\ph)\psi^\pm
      +(\partial_x\ph\pm U(\ph))\psi^\mp],\\
&&\{Q^\pm,Q^\pm\}=2(H \pm (\tilde Z_x - \tilde P_y)),
\eea
and the charge $Q^+$ 
leaves the topological (domain-wall) vacuum $\ph=\ph_K$, $\psi=0$
invariant.
This corresponds to classical
BPS saturation, since with $P_x=0$ and $\tilde P_y=0$
one has $\{Q^+,Q^+\}=2(H+\tilde Z_x)$ and, indeed, with a kink domain wall
$\tilde Z_x/L^{d-1}=W(+v)-W(-v)=-M/L^{d-1}$.

At the quantum level, hermiticity of $Q^\pm$ implies
\be\label{HPyineq}
\< s|H|s\> \ge |\< s|P_y|s\>|  \equiv |\< s|(\tilde P_y-\tilde Z_x)|s\>|.
\ee
This inequality is saturated when
\be
Q^+|s\>=0
\ee
so that BPS states correspond to massless states $P_M P^M=0$
with $P_y=M$ for a kink domain wall in the $x$-direction \cite{Losev:2000mm},
however with infinite momentum and energy unless the $y$-direction
is compact with finite length $L$.
An antikink domain wall has instead $ Q^-|s\>=0$. In both cases,
half of the supersymmetry is spontaneously broken.


Classically, the susy algebra in 1+1 dimensions is obtained from
(\ref{eq:3dsusy}) simply by dropping $\tilde P_y$
as well as $\tilde Z_y$ so that $P_x\equiv\tilde P_x$.
The term $\gamma^2 \tilde Z_x$ remains, however, with
$\gamma^2$ being the nontrivial $\gamma^5$ of 1+1 dimensions.
The susy algebra simplifies to
\be
\{Q^\pm,Q^\pm\} = 2(H\pm Z),\quad \{Q^+,Q^-\}=2P_x
\ee
and one has the inequality
\be
\<s|H|s\> \ge |\<s|Z|s\>|
\ee
for any quantum state $s$. BPS saturated states have
$Q^+|s\>=0$ or $Q^-|s\>=0$, corresponding to
kink and antikink, respectively, and break half of the
supersymmetry.

\section{Quantum corrections to the susy algebra in
dimensional regularization}

\subsection{Fluctuations}

In a kink (or kink domain wall) background one spatial
direction is singled out and we choose this to be along $x$.
The direction orthogonal to the kink direction (parallel
to the domain wall) will be denoted by $y$.

The quantum fields can then be expanded in the
analytically known kink eigenfunctions \cite{Raj:Sol} times plane waves in
the extra dimensions. For the bosonic fluctuations we have
$[-\square+(U{'}^2+UU{''})]\eta=0$ which is solved by
\be
  \label{eq:bosfluct}
\eta=\int\frac{d^{d-1}\ell}{(2\pi)^{\frac{d-1}{2}}}
\intsum\frac{dk}{\sqrt{4\pi\omega}}
    \left(a_{k,\ell}\ e^{-i(\omega t-\ell y)}\phi_k(x)+
      a^{\dagger}_{k,\ell}\ e^{i(\omega t-\ell y)}\phi^{\ast}_k(x)\right).
\ee
The kink eigenfunctions $\phi_k$ are 
normalized according to $\int dx|\phi|^2=1$ for the discrete states 
and to Dirac distributions for the continuum states
according to $\int dx\, \phi_k^*\phi_{k'}=2\pi\delta(k-k')$.
The mode energies are 
$\omega=\sqrt{\omega_k^2+\ell^2}$
where $\omega_k$ is the energy in the 1+1-dimensional case.   

The canonical equal-time commutation relations 
$[\eta(\vec{x}),\dot{\eta}(\vec{x}')]=i\delta(\vec{x}-\vec{x}')$
are fulfilled with
\begin{equation}
  \label{eq:ETC}
  [a_{k,\ell},a^{\dagger}_{k',\ell'}]=\delta_{kk'}\delta(\ell-\ell'),
\end{equation}%
where for the continuum states $\delta_{k,k'}$ becomes a Dirac delta. 

For the fermionic modes which satisfy the Dirac equation $[\ds+U{'}]\psi=0$
one finds
\begin{eqnarray}
  \label{eq:ferm}
 \psi\!&=&\!\psi_0+\int\frac{d^{d-1}\ell}{(2\pi)^{\frac{d-1}{2}}}
\intsum'\frac{dk}{\sqrt{4\pi\omega}}
 \left[b_{k,\ell}\  e^{-i(\omega t-\ell y)}
      { {\scriptstyle\sqrt{\omega+\ell}}\ \phi_k(x)\choose
                    {\scriptstyle\sqrt{\omega-\ell}}\ is_k(x)}
            + b^{\dagger}_{k,\ell}\ (c.c.)\right],\nonumber\\
 &&\psi_0=\int\frac{d^{d-1}\ell}{(2\pi)^{\frac{d-1}{2}}}b_{0,\ell}\ e^{-i\ell (t- y)}
   {\phi_0\choose 0},\quad b^{\dagger}_0(\ell)=b_0(-\ell).
\end{eqnarray}
The fermionic zero mode\footnote{By a slight abuse of notation
we shall always label this by a subscript $0$, but this should not be
confused with the threshold mode $k=0$ (which does not appear
explicitly anywhere below).}
of the susy kink turns into
massless modes located on the domain wall, which have only one
chirality, forming a Majorana-Weyl domain wall fermion 
\cite{Rebhan:2002uk,Callan}.\footnote{The 
mode with $\ell=0$
corresponds in 1+1 dimensions to the zero mode of the susy kink.
It has to be counted as half a degree of freedom in mode
regularization \cite{Goldhaber:2000ab}. For dimensional
regularization such subtleties do not play a role because
the zero mode only gives scaleless integrals and these vanish.}

For the massive modes the Dirac equation relates the eigenfunctions
appearing in
the upper and the lower components of the spinors as follows:
\begin{equation}
  \label{eq:sin}
  s_k=\frac{1}{\omega_k}(\partial_x+U{'})\phi_k
     =\frac{1}{\sqrt{\omega^2-\ell^2}}(\partial_x+U{'})\phi_k,
\end{equation}%
so that the function $s_k$ is the 
SUSY-quantum mechanical \cite{Witten:1982df} partner state of 
$\phi_k$ and thus coincides with the eigen modes of the 
sine-Gordon model (hence the notation) \cite{CooKS:QM}.
With (\ref{eq:sin}), their normalization is the same as that of
the $\phi_k$.

The canonical equal-time anti-commutation relations 
$\{\psi^{\alpha}(\vec{x}),\psi^{\beta}(\vec{x}')\}=\delta^{\alpha\beta}
\delta(\vec{x}-\vec{x}')$ 
are 
satisfied if 
\begin{eqnarray}
  \label{eq:fermetc}
  \{b_0(\ell),b^{\dagger}_0(\ell')\}&=&\{b_0(\ell),b_0(-\ell')\}=\delta(\ell-\ell'),
   \nonumber\\
  \{b_{k,\ell},b^{\dagger}_{k',\ell'}\}&=&\delta_{k,k'}\delta(\ell-\ell'),
\end{eqnarray}%
and again the $\delta_{k,k'}$ becomes a Dirac delta for the continuum states. 
The 
algebra (\ref{eq:fermetc}) 
and the solution for the massless mode (\ref{eq:ferm})
show that the operator $b_0(\ell)$  creates right-moving
massless states on the wall when $\ell$ is negative and 
 annihilates them for positive  momentum $\ell$. 
Thus only massless states with 
momentum in  the positive $y$-direction can be created. 
Changing the representation 
of the gamma matrices by $\gamma^2\rightarrow-\gamma^2$, 
which is inequivalent to 
the original one, reverses the situation. 
Now only massless states with momenta in 
the positive $y$-direction exist. Thus depending on the representation
of the Clifford algebra one chirality of the domain wall fermions
is singled out. This is a reflection of the spontaneous violation 
of parity when embedding the susy kink as a domain wall in 2+1 dimensions.

Notice that in (\ref{eq:ferm}) $d$ can be only 2 or 1, for which
$\ell$ has 1 or 0 components, so for strictly $d=1$ $\ell\equiv0$. 
In order to have a susy-preserving dimensional regularization
scheme by dimensional reduction, we shall start from $d=2$ spatial
dimensions, and then make $d$ continuous and smaller than 2.

\subsection{Energy corrections}

Before turning to a direct calculation of the anomalous
contributions to central charge and momentum, we recapitulate
the one-loop calculation of the energy density of the
susy kink (domain wall) in dimensional regularization.

Expanding the Hamiltonian density of the model (\ref{Lss}),
\begin{equation}
\label{eq:ham}
  \mathcal{H}=\2[\dot{\ph}+(\nabla\ph)^2+{U}^2(\ph)]
   +\2 \psi^{\dagger}i\gamma^0[\vec{\gamma}\ \vec{\nabla}+U'(\ph)]\psi,
\end{equation}%
around the kink/domain wall, using $\ph=\ph_K+\eta$, one obtains
\begin{eqnarray}
  \label{eq:hamexp}
  \mathcal{H}&=&\2[(\partial_x\ph_K)^2+{U}^2]
  -\frac{\delta\mu^2}{\sqrt{2\lambda}}U-\partial_x(U\eta)+{}\nonumber\\
  &&+\2[\dot{\eta}^2+(\nabla\eta)^2+\2(U^2){''}\eta^2]
      +\2\psi^{\dagger}i\gamma^0[\vec{\gamma}\ \vec{\nabla}+U{'}]\psi+
O(\hbar^2),
\end{eqnarray}
where $U$ without an explicit argument implies evaluation at $\ph=\ph_K$
and use of the renormalized $\mu^2$.
The first two terms on the r.h.s. are the classical energy density
and the counterterm contribution. The terms quadratic in the
fluctuations are the only ones contributing to the
total energy.\footnote{The third term in (\ref{eq:hamexp}) is of relevance
when calculating the energy profile \cite{Shifman:1998zy,Goldhaber:2001rp}.} 
They give
\begin{eqnarray}
  \label{eq:h2}
  \int dx\, d^{d-1}y\, \langle\mathcal{H}^{(2)}\rangle&=&
  {L^{d-1}\02}\int dx\int\frac{d^{d-1}\ell}{(2\pi)^{d-1}}
\intsum {dk\02\pi}\Bigl[\frac{\omega}{2}|\phi_k|^2\nonumber\\&&\qquad
  +\frac{1}{2\omega}(\ell^2|\phi_k|^2+|{\phi_k}{'}|^2
   +\2(U^2){''}|\phi_k|^2)\Bigr]\nonumber\\
   &-&{L^{d-1}\02}\int dx\int\frac{d^{d-1}\ell}{(2\pi)^{d-1}}
\intsum {dk\02\pi}\frac{\omega}{2}(|\phi_k|^2
   +|s_k|^2),\quad
\end{eqnarray}
where the two sum-integrals are the bosonic and fermionic
contributions, respectively.

Using $\2(U^2){''}=U{'}^2-\partial_xU{'}$ which follows from
the Bogomolnyi equation $\6_x\ph_K=-U$ and partially integrating
(or alternatively from the equipartition theorem for the energy of the
bosonic fluctuations in (\ref{eq:hamexp})),
one obtains
the expected sum-integrals over zero-point energies,
\bea\label{zeropoints}
\int dx\, d^{d-1}y\, \langle\mathcal{H}^{(2)}\rangle =
{L^{d-1}\02}\int dx\int\frac{d^{d-1}\ell}{(2\pi)^{d-1}}
\intsum {dk\02\pi}\frac{\omega}{2}(|\phi_k|^2
   -|s_k|^2).
\eea
In these expressions, the massless modes (which correspond to the zero mode
of the 1+1 dimensional kink) can be dropped in dimensional regularization
as scaleless and thus vanishing contributions, and the massive discrete
modes cancel between bosons and fermions.\footnote{The zero mode
contributions in fact do not cancel by themselves
between bosons and fermions, because the
latter are chiral. This noncancellation is in fact crucial in energy cutoff
regularization (see Ref.~\cite{Rebhan:2002uk}).}
Carrying out the $x$-integration over the continuous mode functions
gives a difference of spectral densities, namely
\be\label{specdiff}
\int dx (|\phi_k(x)|^2-|s_k(x)|^2)=-\theta'(k)
=-{2m\0k^2+m^2},
\ee
where $\theta(k)$ is the additional phase shift of the mode functions
$s_k$ compared to $\phi_k$.

Combining (\ref{zeropoints}) and (\ref{specdiff}), and adding
in the counterterm contribution from (\ref{deltamu2})
leads to a simple integral
\bea\label{Mkink}
{\Delta M^{(1)}\0L^{d-1}}&=&-{1\04}\int {dk\,d^{d-1}\ell\0(2\pi)^d}
{\omega}\,\theta'(k)+m\delta v^2\nonumber\\
&=&-{1\04}\int {dk\,d^{d-1}\ell\0(2\pi)^d}{\ell^2\0\omega}\theta'(k)
=-{2\0d}{\Gamma({3-d\02})\0(4\pi)^{d+1\02}}\, m^d.
\eea
This reproduces the correct known result for the susy kink mass
correction $\Delta M^{(1)}=-m/(2\pi)$ (for $d=1$) and
the surface (string) tension of the
2+1 dimensional susy kink domain wall 
$\Delta M^{(1)}/L=-m^2/(8\pi)$ (for $d=2$) 
\cite{Rebhan:2002uk}.

Notice that the entire result is produced by an integrand
proportional to the extra momentum component $\ell^2$,
which for strictly $d=1$ would not exist.
This can also be observed by recasting $\langle\mathcal{H}^{(2)}\rangle$
in (\ref{eq:h2}) with the help of (\ref{eq:sin}) in the form
\begin{eqnarray}
  \label{eq:h22}
   \langle\mathcal{H}^{(2)}\rangle&=&-\partial_x
 \left(\2\int\frac{d^{d-1}\ell}{(2\pi)^{d-1}}\intsum {dk\02\pi} \ 
   U{'}\frac{|\phi_k|^2}{2\omega}\right)+{}\nonumber\\
 &&{}+\2\int\frac{d^{d-1}\ell}{(2\pi)^{d-1}}\intsum {dk\02\pi}
   \frac{\ell^2}{2\omega}(|\phi_k^2|-|s_k|^2).
\end{eqnarray}%
When integrated, the first term, which is a pure surface term,
cancels exactly the counterterm (see (\ref{deltamu2})), because
\begin{eqnarray}
  \label{eq:surface}
  \int dx\langle\2\partial_x(U{'}\eta^2)\rangle=
    \2 U{'}\langle\eta^2\rangle|_{-\infty}^{\infty}=
    m\int\frac{d^{d-1}\ell}{(2\pi)^{d-1}}\int\frac{dk}{2\pi}\frac{1}{2\omega}
\equiv m \delta v^2,
\end{eqnarray}%
where we have used that $U{'}(x=\pm\infty)=\pm m$.
The second contribution in (\ref{eq:h22}), on the other hand,
is precisely the r.h.s. of (\ref{Mkink}).

\subsection{Anomalous contributions to the central charge and
extra momentum}
\label{sec33}

In a kink (domain wall) background with only nontrivial $x$ dependence,
the central charge density $\tilde \mathcal Z_x$ receives
nontrivial contributions.
Expanding $\tilde \mathcal Z_x$ 
around the kink background gives
\begin{eqnarray}
  \label{eq:Znaiv}
  \tilde{\mathcal{Z}}_x=U\partial_x\ph_K-\frac{\delta\mu^2}{\sqrt{2\lambda}}
        \partial_x\ph_K+\partial_x(U\eta)+\2\partial_x(U{'}\eta^2)
        +O(\eta^3). 
\end{eqnarray}
Again only the part quadratic in the fluctuations contributes to
the integrated quantity at one-loop order\footnote{Again, this does not hold
for the central charge density locally 
\cite{Shifman:1998zy,Goldhaber:2001rp}.}.
However, this leads just to the contribution shown in (\ref{eq:surface}),
which matches
precisely the counterterm $m\delta v^2$ from requiring vanishing tadpoles.
Straightforward application of the rules of
dimensional regularization thus leads to a null result for
the net one-loop correction to $\<\tilde Z_x\>$ in the same way
as found in Refs.~\cite{Imbimbo:1984nq,Rebhan:1997iv,Nastase:1998sy}
in other schemes.

On the other hand, by considering the less singular
combination $\<H+\tilde Z_x\>$ and showing that it vanishes exactly, 
it was concluded in Ref.~\cite{Graham:1998qq}
that $\<\tilde Z_x\>$ has to compensate any nontrivial result
for $\<H\>$, which in Ref.~\cite{Graham:1998qq} was obtained
by subtracting successive Born approximations for scattering
phase shifts. In fact, Ref.~\cite{Graham:1998qq} explicitly
demonstrates how to rewrite $\<\tilde Z_x\>$ into $-\<H\>$,
apparently without the need for the anomalous terms in the quantum central
charge operator
postulated in Ref.~\cite{Shifman:1998zy}.

The resolution of this discrepancy is that Ref.~\cite{Graham:1998qq}
did not regularize $\<\tilde Z_x\>$ and the manipulations
needed to rewrite it as $-\<H\>$ (which eventually is
regularized and renormalized) are ill-defined.
Using dimensional regularization one in fact obtains
a nonzero result for $\<H+\tilde Z_x\>$, apparently
in violation of susy.

However, dimensional regularization by embedding the kink
as a domain wall in (up to) one higher dimension, which
preserves susy, instead leads to 
\be
\<H+\tilde Z_x-\tilde P_y\>=0,
\ee
i.e. the saturation of (\ref{HPyineq}), as we shall now verify.

The bosonic contribution to $\<\tilde P_y\>$ involves
\begin{equation}
  \label{eq:pybos}
  \2\langle\dot{\eta}\partial_y\eta+\partial_y\eta\dot{\eta}\rangle=
  -\int\frac{d^{d-1}\ell}{(2\pi)^{d-1}}\intsum {dk\02\pi}\ 
\frac{\ell}{2}|\phi_k(x)|^2.
\end{equation}%
The $\ell$-integral factorizes and gives zero both because it is
a scale-less integral and because the integrand is odd in $\ell$.
Only the fermions turn out to give interesting contributions:
\begin{eqnarray}
  \label{eq:py}
  \langle\tilde{\mathcal{P}}_y\rangle&=&
  \frac{i}{2}\langle\psi^{\dagger}\partial_y\psi\rangle\nonumber\\
&=& \2 \int\frac{d^{d-1}\ell}{(2\pi)^{d-1}}\intsum {dk\02\pi}{\ell\02\omega}
 \left[(\omega+\ell)|\phi_k|^2+(\omega-\ell)|s_k|^2\right]\nonumber\\
  &=&\2\int\frac{d^{d-1}\ell}{(2\pi)^{d-1}}\ \ell\ \theta(-\ell)\ |\phi_0|^2+
              \nonumber\\
  &&+\2\int\frac{d^{d-1}\ell}{(2\pi)^{d-1}}\intsum' {dk\02\pi}
   \left( \frac{\ell}{2}(|\phi_k|^2+|s_k|^2)
    +\frac{\ell^2}{2\omega}(|\phi_k|^2-|s_k|^2)
\right).\qquad
\end{eqnarray}
{}From the last sum-integral we have separated off the contribution
of the zero mode of the kink, which turns into chiral domain wall fermions
for $d>1$. The contribution of the latter no longer vanishes by
symmetry, but the $\ell$-integral is still scale-less and therefore
put to zero in dimensional regularization. The first sum-integral
on the right-hand side is again zero by both symmetry and scalelessness,
but the final term is not. The $\ell$-integration no longer
factorizes because $\omega=\sqrt{k^2+\ell^2+m^2}$, and
is in fact identical to the finite contribution
in $\<\mathcal H\>$ obtained already in (\ref{eq:h22}).

So for all $d\le2$ we have BPS saturation, $\<H\>=|\<\tilde Z_x-\tilde P_y\>|$,
which in the limit $d\to1$, the susy kink, is made possible by 
a nonvanishing $\<\tilde P_y\>$. The anomaly in the central charge
is seen to arise from a parity-violating contribution in $d=1+\epsilon$
dimensions which is the price to be paid for preserving supersymmetry
when going up in dimensions to embed the susy kink as a domain wall.

It is perhaps worth emphasizing that the above results do not
depend on the details of the spectral densities associated with
the mode functions $\phi_k$ and $s_k$. In the integrated quantities
$\<H\>$ and $\<\tilde P_y\>$ only the difference of the
spectral densities as given by (\ref{specdiff}) is responsible for
the nonvanishing contribution. The function $\theta(k)$ therein is
entirely fixed by the form of the Dirac equation in the asymptotic
regions $x\to\pm\infty$ far away from the kink \cite{Rebhan:1997iv}.

\section{Dimensional reduction and evanescent counterterms}

In the above, we have effectively used the 't Hooft-Veltman version
of dimensional regularization \cite{'tHooft:1972fi}
in which the space-time dimensionality $n$
is made larger than the dimension of interest.
In general this breaks susy because the numbers of bosons and fermions
are not the same anymore when one moves up in dimensions. But in our
particular model the number of states are the same in 1+1 and 2+1
dimensions, so that we could preserve susy, though this led to
new physics like spontaneous parity violation and chiral domain wall fermions.

In 2+1 dimensions, we have $P_y=\tilde P_y-\tilde Z_x$
and $|\<P_y\>|=\<H\>$, where $\tilde P$ and $\tilde Z$ were
defined in (\ref{eq:3dsusy}). Classically, this BPS saturation
is guaranteed by $\tilde Z_x$ alone. At the quantum level,
however, the quantum corrections to the latter are cancelled
completely by the counterterm from renormalizing tadpoles to zero.
All nontrivial corrections come from the ``genuine'' momentum
operator $\tilde P_y$, and are due to having a spontaneous
breaking of parity.

In the limit of 1+1 dimensions, because $\gamma^2|_{D=2+1}=
\gamma^5|_{D=1+1}$, one has to make the
identification $Z=\tilde Z_x-\tilde P_y$. For $\tilde Z_x$, one
again does not obtain net quantum corrections. However, the
expectation value $\<\tilde P_y\>$ does not vanish in the
limit $d\to1$, although there is no longer an extra dimension.
The spontaneous parity violation in the 2+1 dimensional theory, which had to be
considered in order to preserve susy, leaves a finite imprint upon
dimensional reduction to 1+1 dimensions by providing
an anomalous additional contribution to $\<\tilde Z_x\>$ balancing
the nontrivial quantum correction $\<H\>$. 

We now comment on how the central charge anomaly can be recovered
from Siegel's version of dimensional regularization 
\cite{Siegel}
where $n$ is smaller than the dimension of spacetime and where one keeps
the number of field components fixed, but lowers the number of
coordinates and momenta from 2 to $n<2$. At the one-loop level one
encounters 2-dimensional $\delta_\mu^\nu$ coming from
Dirac matrices, and $n$-dimensional $\hat\delta_\mu^\nu$ from
loop momenta. An important concept which is going to play a role are
the evanescent counter\-terms \cite{Bonneau}
involving the factor ${1\0\epsilon}\hat{\hat{\delta}}{}_\mu^\nu
\gamma_\nu\psi$, where $\hat{\hat\delta}{}_\mu^\nu\equiv \delta_\mu^\nu-
\hat\delta_\mu^\nu$ has only $\epsilon=2-n$ nonvanishing components.

For the chiral anomaly in 3+1 dimensions due to a massless Dirac fermion
coupled to on-shell photons one finds from dimensional reduction
the following expression for the regularized but not yet
renormalized chiral current~\cite{Grisaru}
\be
j_\mu=\2\6_\mu{1\0\Box}F^{\rho\sigma}\tilde F_{\rho\sigma}-{2\0\epsilon}
\tilde F_{\mu\nu} \hat{\hat A\,}\!{}^\nu
\ee
where $\tilde F_{\mu\nu}=\2\epsilon_{\mu\nu\rho\sigma}F^{\rho\sigma}$
and $\hat{\hat A\,}\!{}^\nu \equiv \hat{\hat\delta}{}_\mu^\nu A^\mu$.
This current is gauge invariant because $\delta\hat{\hat A\,}\!{}_\nu=
\hat{\hat\partial\,}\!{}_\nu\lambda=0$ as coordinates only lie in the
$n$-dimensional subspace. It is conserved since total antisymmetrization
of five indices in 4 dimensions yields
\be
\tilde F^{\mu\nu} \6_\mu \hat{\hat A\,}\!{}_\nu=
\2\epsilon^{\mu\nu\rho\sigma}F_{\rho\sigma}\6_\mu \hat{\hat\delta}{}_\nu^\lambda A_\lambda
=-\tilde F^{\rho\nu}\6_\rho \hat{\hat A\,}\!{}_\nu+\epsilon\2
F^{\mu\nu}\tilde F_{\mu\nu}.
\ee
Clearly, $\6^\mu j_\mu=\2 F^{\mu\nu}\tilde F_{\mu\nu}-
{2\0\epsilon}(\4\epsilon F^{\mu\nu}\tilde F_{\mu\nu}) = 0$.
The composite operator $j_\mu$ is renormalized by subtracting the
divergence $-{2\0\epsilon}\tilde F_{\mu\nu}\hat{\hat A\,}\!{}^\nu$
(operator mixing),
and thus in dimensional reduction the chiral anomaly is produced by the
(evanescent) counterterm, and not by the loop graph.

Consider now the supercurrent 
$j_\mu=-(\not\!\partial\ph+U(\ph))\gamma_\mu\psi$.
In the trivial vacuum, expanding into quantum fields yields
\be\label{jmuexp}
j_\mu=-\left(\not\!\partial\eta+U'(v)\,\eta
+
\2 U''(v)\,\eta^2
\right)\gamma_\mu
\psi + {1\0\sqrt{2\lambda}}\delta\mu^2\gamma_\mu\psi.
\ee
Only matrix elements with one external fermion are divergent.
The term involving $U''(v)\eta^2$ in (\ref{jmuexp}) gives rise to
a divergent scalar tadpole that is cancelled 
completely by the counterterm $\delta\mu^2$ (which
itself is due to an $\eta$ and a $\psi$ loop). The only other
divergent diagram is due to the term involving $\not\!\partial\eta$
in (\ref{jmuexp}) and has the form a $\psi$-selfenergy. Its
singular part reads
\be
\<0| j_\mu |p\>^{\rm div} = i U''(v)
\int_0^1 dx \int {d^n \kappa\0(2\pi)^n} {\not\!\kappa \gamma_\mu\! \not\!\kappa
\over [\kappa^2 + p^2 x(1-x) + m^2]^2}u(p).
\ee
Using $\hat\delta_\mu^\nu\equiv \delta_\mu^\nu-\hat{\hat\delta}{}_\mu^\nu$
we find that under the integral
$$\not\!\kappa \gamma_\mu\!\! \not\!\kappa = - \kappa^2(\delta_\mu^\lambda-
{2\0n}\hat \delta_\mu^\lambda)\gamma_\lambda=
{\epsilon\0n}\kappa^2\gamma_\mu - {2\0n}\kappa^2
\hat{\hat\delta}{}_\mu^\lambda \gamma_\lambda$$
so that
\be
\<0| j_\mu |p\>^{\rm div} = {U''(v)\02\pi}{\hat{\hat\delta}{}_\mu^\lambda
\0\epsilon} \gamma_\lambda u(p).
\ee
Hence, the regularized one-loop contribution to the susy 
current contains the evanescent operator
\be
j_\mu ^{\rm div} = {U''(\ph)\02\pi}{\hat{\hat\delta}{}_\mu^\lambda
\0\epsilon} \gamma_\lambda  \psi.
\ee
This is by itself a conserved quantity, because all fields
depend only on the $n$-dimensional coordinates, but it
has a nonvanishing contraction with $\gamma^\mu$.
The latter 
gives rise to an anomalous contribution to the 
renormalized conformal-susy current
$\not \!x j_\mu^{\rm ren.}$ where $j_\mu^{\rm ren.}=j_\mu-j_\mu^{div}$,
\be
\partial^\mu (\not\!x j_\mu^{\rm ren.})_{\rm anom.}=-
\gamma^\mu j_\mu^{\rm div}= -{U''\02\pi}\psi.
\ee
(There are also nonvanishing
nonanomalous contributions
to $\partial^\mu (\not\!x j_\mu)$ because our model is not
conformal-susy invariant at the classical level.)

Ordinary susy on the other hand is unbroken; there is no anomaly
in the divergence of $j_\mu^{\rm ren.}$. A susy variation of
$j_\mu$ involves the energy-momentum tensor and the
topological central-charge current $\zeta_\mu$
according to
\be
\delta j_\mu = -2T_\mu{}^\nu \gamma_\nu \epsilon - 2 \zeta_\mu \gamma^5 
\epsilon,
\ee
where classically $\zeta_\mu=\epsilon_{\mu\nu}U
\partial^\nu \varphi$.

At the quantum level,
the counter-term $j_\mu^{\rm ct}=-j_\mu^{\rm div.}$ induces
an additional contribution to the central charge current
\be
\zeta_\mu^{\rm anom}={1\04\pi}
{\hat{\hat\delta}{}_\mu^\nu\0\epsilon}
\epsilon_{\nu\rho}\partial^\rho U'
\ee
which despite appearances is a {\em finite} quantity: using that total
antisymmetrization of the three lower indices has to vanish
in two dimensions gives
\be
\hat{\hat\delta}{}_\mu^\nu
\epsilon_{\nu\rho} = \epsilon \epsilon_{\mu\rho}+
\hat{\hat\delta}{}_\rho^\nu
\epsilon_{\nu\mu} 
\ee
and together with the fact the $U'$ only depends on $n$-dimensional
coordinates this finally yields
\be\label{zetaanom}
\zeta_\mu^{\rm anom}={1\04\pi} \epsilon_{\mu\rho} \partial^\rho U'
\ee
in agreement with the anomaly in the central charge as obtained
previously.

We emphasize that $\zeta_\mu$ itself does not require the
subtraction of an evanescent counterterm. The latter only
appears in the susy current $j_\mu$, which gives rise
to a conformal-susy anomaly in $\not \!x j_\mu$.
A susy variation of the latter
shows that it forms a conformal current multiplet involving besides
the dilatation current $T_{\mu\nu}x^\nu$ and the Lorentz current
$T_\mu{}^\nu x^\rho\epsilon_{\nu\rho}$
also a current 
$j_{(\nu)}^{(\zeta)\mu}=x^\rho \epsilon_{\rho\nu}\zeta^\mu$.
We identify this with the conformal central-charge current, which
is to be distinguished from the ordinary central-charge current $\zeta_\mu$.

Since $\partial_\mu j_{(\nu)}^{(\zeta)\mu}=\epsilon_{\mu\nu}\zeta^\mu$,
and $\epsilon_{\mu\nu}$ is invertible, the entire
central-charge current $\zeta^\mu$ enters in the divergence
of the conformal central-charge current,
whereas in the case of the conformal-susy current it was 
the contraction $\gamma_\mu j^\mu$.

The current $j^{(\zeta)}$ thus has the curious property of being
completely determined by its own divergence. For this reason it 
is in fact not associated with any continuous symmetry
(as is also the case for the ordinary central-charge current,
which is of topological origin). In the absence
of classical breaking of conformal invariance it is conserved
trivially by its complete disappearance and then there is no
symmetry generating charge operator. Nevertheless, in the 
conformally noninvariant susy kink model this current arises
and has in addition to its nonanomalous divergence an 
anomalous one, namely the
anomalous contribution to the central charge current
inherited from the evanescent counterterm in the renormalized susy current.

\small


\end{document}